\documentclass[12pt,showpacs,showkeys,english]{revtex4}
\usepackage{subfigure}
\usepackage{graphicx}
\usepackage{amssymb}
\usepackage{color}

\makeatletter

\usepackage{babel}
\makeatother

\begin{document}

\title{
Coupled Intermittent Maps Modelling the Statistics
of Genomic Sequences: A Network Approach}

\author{Astero Provata$^{1,2}$ and Christian Beck$^{2}$}

\affiliation{$^1$Laboratory of Statistical Mechanics and
Nonlinear Dynamics, National Center for Scientific
Research ``Demokritos\char`\"{}, 15310 Athens, Greece \\
$^2$School of Mathematical Sciences, Queen Mary, University of
London, Mile End Road, London E1 4NS,UK}

\begin{abstract}

The dynamics of coupled intermittent maps is used to model the correlated
structure of genomic sequences. The use of intermittent
maps, as opposed to other simple chaotic maps, is particularly suited
for the production of long range correlation features which are observed
in the genomic sequences of higher eucaryotes.
A weighted network approach to symbolic sequences
is introduced and it is shown that coupled intermittent polynomial
maps produce degree and link size
distributions with power law exponents
similar to the ones observed
in real genomes.
The proposed network approach to symbolic sequences is
generic and can be applied to any symbol sequence (artificial or natural).

\end{abstract}


\date{\today}

\maketitle

\section{Introduction}
\label{sec1}
\par Some 20 years ago, in 1992, the presence of long range
correlations in genomic sequences was first reported
 in three seminal papers \cite{li:1992,peng:1992,voss:1992}.
Since then many attempts were made to record, classify and  model
these genomic correlations
 and to connect them with the functionality and evolution
of the current day genome
\cite{li:1992,peng:1992,voss:1992,arneodo:2010,buldyrev:1995,herzel:1998,li:1994,katsaloulis:2006,messer:2007}.
Despite these many attempts a conclusive explanation of the
presence and the role of long range correlations in the genome
is still missing.

\par In an earlier publication \cite{provata:2010},
one of the current authors (A.P.)
and P. Katsaloulis have searched for a hierarchical process
which could produce long range correlations similar to the
ones observed in genomic sequences. To this end, they
introduced a 2D density correlation  matrix M
which is based on the frequency of appearance of blocks/strings
of size $s$. They calculated the multifractal
properties of DNA from this matrix and from its multiple
superpositions to create strings of longer lengths.
In fact, their approach
corresponds to a description of all strings of multiple lengths
$2s,4s\cdots$, assuming that
the correlations are negligible for length scales $l > s$.
This method produces correlations up to finite
scales, which are comparable with the ones observed in the
genome at the same length scales \cite{provata:2010}. Nevertheless, long-range correlations
are known to persist over many scales in DNA and are not
limited to a finite length scale \cite{provata:2007,katsaloulis:2009}.
In a further quest for dynamical mechanisms producing long
range correlations over extended scales
the current study uses the dynamics of intermittent
maps to produce symbol sequences with characteristics similar
to DNA.
\par Intermittent maps are well-known
to produce a variety of interesting features such as
metastable behaviour and anomalous transport, often
characterised by long-term correlations and power laws \cite{dettmann:1998,
balakrishnan:2001,
korabel:2005, korabel:2007,pollicott:2009, froyland:2011, bhansali:2005}.
That is why they are particularly suited for the modelling of the dynamics
of DNA strands with long range features, such as the genome of higher
eucaryotes. In particular, the polynomial map\cite{bhansali:2007eco}
is particularly suited for the DNA modeling due to its simplicity,
versatility and the large parameter range which gives rise to long range
characteristics. This map will be used in the modeling of genomic data
by first transforming the times series generated by the map
into a symbol sequence and then comparing
its statistics with that of whole eucaryotic chromosomes.

\par For the comparison between the dynamics produced by the
intermittent polynomial map and that of genomic sequences
a novel network approach  will first be established.
For this, the time series produced by the polynomial maps
will be transformed into symbol sequences and then associated
networks will be constructed. The properties of these networks
(degree distribution, link size distribution, clustering
coefficients) will be computed both for the polynomial map
and for the genomic sequences and the statistics will be compared. It will
turn out that the use of single polynomial maps is not
enough to produce the exact power law exponents observed in
the network description of the genome. The solution to this
problem is given by weakly coupling the polynomial maps on a lattice.
The weak coupling modifies the power law exponents of the
zero-coupling limit and produces power law tails comparable
to the ones observed for the genome data.

\par This work has the following structure:
In the next section the dynamics of the intermittent polynomial map
is briefly recapitulated and the corresponding symbol sequence
is constructed. The construction of a dynamical
weighted network method for the description of correlations in
symbol sequences is presented in the same section.
  In Sec. \ref{sec3} the network method is applied to both,
human chromosomes and symbol sequence of the intermittent
polynomial map. Comparative results are presented and discussed.
In Sec. \ref{sec4}, coupled
polynomial maps are discussed. It is shown that small 
couplings give exponents very close to the ones observed
for genomic sequences.
In our concluding remarks of section \ref{sec5} the general use of
the network method is summarized.

\section{Intermittent Maps and Associated Networks}
\label{sec2}
In this section we first recall the dynamics of the polynomial
map and describe the transformation to symbol sequence for later comparison
with genomic sequences. It is important to
note here that uniformly distributed symbol frequencies will not be assumed
in the current study. The symbol frequency produced by the map
will depend on the chosen partition of the
phase space and will be dictated by comparison with real genomic sequences
where the symbol frequencies have different average values for each symbol.

\subsection{The Polynomial Map}
\label{polynomial-map}
\begin{figure}
\includegraphics[clip,width=0.6\textwidth,angle=0]{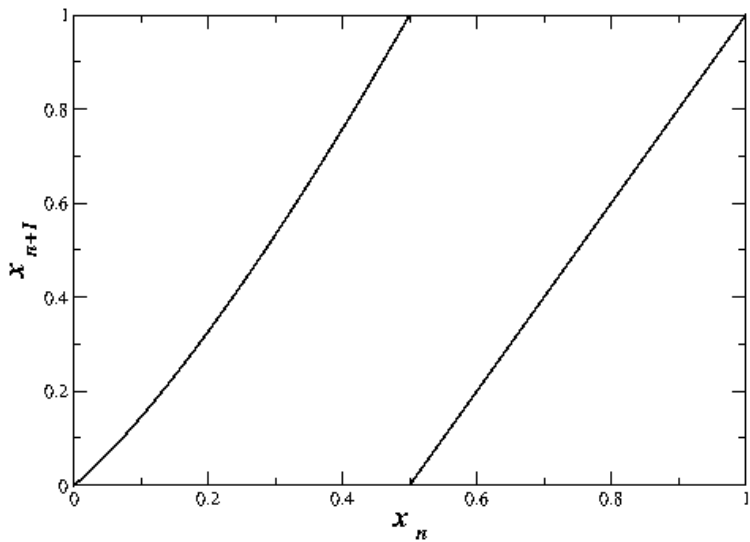}
\caption{\label{fig:01}{} (Colour online)
The polynomial map with parameter value
$\alpha =0.5$.
}
\end{figure}

The polynomial map is defined by the following iteration
scheme \cite{bhansali:2007,bhansali:2005}
\begin{eqnarray}
x_{n+1}= \left\{
\begin{array}{l l l}
x_{n}(1+2^{\alpha}x_{n}^{\alpha }),&\quad {\rm if} & x_{n}\le 0.5 \\
2x_{n},&\quad {\rm if} & x_{n}>0.5 \\
\end{array}
\right.
 \>\> n=1,2...
\label{eq01}
\end{eqnarray}
where $n$ is a discrete time index and $x_n \in [0,1]$ is taken modulo 1 for all $n$
and $\alpha >0$. For $0< \alpha <1$ the map is ergodic. Figure
\ref{fig:01} shows the graph of $x_{n+1}$ vs. $x_n$
 of the polynomial map for a parameter $\alpha=0.5$ located in the center of the
ergodic regime.
Note that near $x_n\sim 0$ the slope is close to 1 and hence
intermittent behavior is produced. In a symbolic dynamics approach,
the laminar phase of intermittent behavior corresponds to
repetitions of the same symbol for quite a long time,
which is then interrupted by chaotic outbursts \cite{BS}.
The symbol repetitions generate long-term correlations. A similar
feature is also observed
in genomic sequences. For DNA very often particular substrings of
symbols are repeated again and again, causing a dynamics that
is significantly different from random behaviour and
exhibiting long-term correlations. For this reason it is
obvious that intermittent maps,
as opposed to simple, fully developed chaotic maps, are good candidates to
model sequences of symbols with similar statistics as in genomes.

\subsection{Symbol Sequences Associated with Maps}
\label{symbol-sequences}
The symbolic dynamics technique for the analysis of maps
has a long tradition (see, e.g., \cite{BS} for an introduction).
The resulting sequences
carry the correlations inherited by the map and provide the means of
understanding the dynamical behavior in a coarse-grained way.

\par To comply with the structure of
genomic sequences we use a translation based on $m=4$ symbols.
The phase space is partitioned into four
segments $[ 0,M_1) ,\>
[ M_1,M_2) ,\> [ M_2,M_3) ,\> [ M_3,1 ] $, where $M_i,\, i=1,2,3$
are real numbers, chosen in such a way
that the frequency of appearance of the four nucleotides
in a particular chromosome is reproduced by the map.
Using this phase space partition the time series produced 
by Eq. \ref{eq01} is transformed into a
sequence $L=l_1,l_2,l_3, \ldots$, with symbols taken
from a 4-letter alphabet representing
 the four nucleotides: $l_i\in [ A(Adenine),G(Guanine),
C(Cytosine),T(Thymine)] $.
\begin{eqnarray}
l_{i}= \left\{
\begin{array}{l l l}
A,&\quad {\rm if} & 0\le x_i<M_1 \\
G,&\quad {\rm if} & M_1\le x_i<M_2 \\
C,&\quad {\rm if} & M_2\le x_i<M_3 \\
T,&\quad {\rm if} & M_3\le x_i<1
\end{array}
\right.
 \>\> i=1,2...
\label{eq02}
\end{eqnarray}

As a particular example we consider Human Chromosome 20, where
the individual nucleotide frequencies are:
$p_A=0.282856, p_C=0.215134,p_G=0.215896$ and $p_T=0.286114$.
To calculate the $M_i$ values, we first 
determine the invariant density of
the polynomial map, i.e. we iterate the map and calculate the
local density of points, or probability $p(x)$ that a specific value
will occur between $x$ and $x+dx$.
For this chromosome the $M_i$ $i=1,\cdots 4$, are determined as:
\begin{eqnarray}
\begin{array}{l}
\int_0^{M_1}p(x) dx=p_A=0.282856 \\
\int_{M_1}^{M_2}p(x) dx=p_G=0.215896 \\
\int_{M_2}^{M_3}p(x) dx=p_C=0.215134
\end{array}
\label{eq03}
\end{eqnarray}

By using the transformation Eq. \ref{eq02} of map Eq. \ref{eq01} with
$M_j$ values given by Eq. \ref{eq03} an arbitrarily
long symbol sequence $l_i, i=1,\cdots, N$ is produced, whose correlations
are dictated by the polynomial map and whose symbol frequencies correspond
to the ones of chromosome 20.

\subsection{Network Approach to Symbolic Sequences}
\label{networks}
In this section a general relation between networks and
symbol sequences is established.
This construction is generic and holds for any symbol sequence whether it
is a natural or experimental symbol sequence (eg. natural languages, DNA)
or an artificial sequence. In the second category random sequences are
included, as well as sequences obtained via certain rules/algorithms and sequences
obtained e.g.\ by map iteration processes, as described in the previous section.

\par Consider a generic symbolic sequence
\begin{eqnarray}
L=l_1,l_2,\cdots l_i\cdots l_N
\label{eq04}
\end{eqnarray}
 of length $N$, where the symbols
$l_i$ take values from a finite alphabet of size $m$.
 For our approach the sequence $L$ is covered
with (divided into) segments (blocks, strings) of size $s << N$.
The maximum number of all possible strings of size $s$
with symbols taken from an alphabet with $m$ symbols is

\begin{eqnarray}
S_{max}=m^s
\label{eq05}
\end{eqnarray}

To fully cover the sequence, $N/s$ segments are needed .
 As a concrete example
consider covering the binary ($m=2$) sequence $L=\{ 001010001101011011001\}$
by strings of size $s=3$. The following substrings occur:
 $ S_1=\{ 001\} ,\>\> S_2=\{ 010\} ,\>\> S_3=\{  101\} ,\>\>
S_4=\{ 011\}$ with string $S_1$ occurring three times and string $S_4$
occurring twice.
\par In uncorrelated, random sequences of infinite size
all strings $S_i, \>\> i=1,\cdots
S_{max}$ of length $s$ occur with the same probability
\begin{eqnarray}
p_i=1/S_{max},\>\> i=1,\cdots S_{max}
\label{eq06}
\end{eqnarray}
while for correlated and natural sequences Eq. \ref{eq06}
usually does not hold. In natural and correlated sequences the total
number of observed strings is denoted by $V$ and is always $V\le S_{max}$.
\par Within the ensembles of possible $S_i$ consider
furthermore the probability $b_{ij}$
of string $i=[ I_1,I_2\cdots I_s ]$ to be followed
by string $j=[ J_1,J_2\cdots J_s ]$
(both having the same length $s$).  The elements $b_{ij}$ are identified
actually as conditional probabilities: having located the
string $i$ in the sequence $L$, the element $b_{ij}$ represents the
conditional probability that it is followed by the string $j$.
$\mathbf{b}$ is a square matrix of size $V \times V $.
The matrix $\mathbf{b}$ can be related to the
joint probability of finding the combined string
$i\otimes j= [ I_1,I_2\cdots I_s, J_1,J_2\cdots J_s ]$ of length $2s$
as follows:

\begin{eqnarray}
b_{ij}=\frac{p_{i\otimes j}}{p_i}=
\frac{p_{[ I_1,I_2\cdots I_s, J_1,J_2\cdots J_s]}}{p_{[I_1, \cdots , I_s]}}
\label{eq07}
\end{eqnarray}

\par Based on the conditional probability $b_{ij}$ of string $i$
to be followed by string $j$
on a very long sequence $L$, an associated,  abstract
network can be constructed whose nodes are the strings
 $S_i, \>\> i=1,\cdots V $ of length $s$.
Thus the number of nodes, or {\it network capacity} $V$,
is at most $S_{max}$.
An edge is drawn
between two nodes $i$ and
$j$ if the corresponding strings $i=[ I_1,I_2\cdots I_s ]$
and $j=[ J_1,J_2\cdots J_s ]$
are found in direct succession anywhere in the sequence $L$.
The edge between
$i$ and $j$ nodes is weighted with the frequency of finding
strings $i$ and $j$
in succession and thus the conditional probability matrix element $b_{ij}$
gives the weight of the edge between nodes $i$ and $j$.
In the network notation the matrix $b_{ij}$ is
identified as the {\it connectivity} or {\it adjacency} matrix.
Note that in general $b_{ij}\neq b_{ji}$ for genomic or natural symbolic sequences.
Thus the adjacency matrix created by genomic sequences indicates that
the corresponding network belongs to the class of
{\it directed} networks/graphs.

\par
In the abstract networks generated by symbolic sequences as proposed above, loops
(sometimes also called
''self-loops'' or ''buckles'') are often present, since it is
quite common that
a certain string will be followed by an identical string.
Loops do not occur in social networks, for example, where
an individual does not interact with himself. On the other hand, in
food distribution networks between cities self-loops on nodes are allowed,
since food maybe consumed (or distributed) in the city it was produced.
Loops are also observed in genomic networks,
brain neuron networks, cardio-vascular
system etc.
\cite{kivela:2004,makitie:1999,ahn:2009,dojer:2006}.
In terms of the elements of the connectivity matrix, the presence
of loops means $b_{ii}\neq 0$. In graph theory, graphs
which contain loops are often called {\it multigraphs}.
\par

Having defined the nodes and links in the network corresponding to
a symbol sequence we proceed in identifying the various network
parameters. The {\it degree} $k_i$ of a node $i$, which corresponds to
the symbolic string $i=[I_1,I_2,\cdots I_s ]$,  is usually defined as
the number of links originating from the
node $i$ towards any other node in the system.
For weighted networks, as in the case of symbol
sequences, each link is weighted with the appropriate
weighting factor and the degree $k_i$ expresses the
cumulative weighted linking of the particular node $i$ to
all other network nodes. In the case of symbol sequences,
(where the links are identified as the conditional probabilities $b_{ij}$),
the outflowing degree $k_i$ of string $i$ is calculated as
\begin{eqnarray}
k_i=\sum_{j=1}^Vb_{ij}=\frac{\sum_{j=1}^Vp_{i\otimes j}}{p_i}=1
\label{eq07777}
\end{eqnarray}

Thus, when we use the conditional probability $b_{ij}$, all nodes
carry the same outflowing degree (normalized to 1), since each string is always followed by another
string within the $V$ possible strings.  
However, since we are dealing with directed networks, we also have
to take into account the inflowing degrees of freedom. The probability
to observe a certain string $i$ is then given by the balance between
inflow and outflow.

 In the case of symbol sequences we identify
the degree $k_i$ of a node $i$ as the frequency of
appearance of the corresponding string $i$, to be consistent 
with the distributed weights carried by the nodes .
This definition makes sense: For dynamical systems
with a Markov partition the invariant probabilities of string sequences are determined by the balance between inflowing and
outflowing iterates (a direct consequence of the
fixed-point property of the Perron-Frobenius operator). Hence
the net balance of flow along the links fixes the invariant density
and hence also the probabilities of symbol sequences in a coarse-grained
description.

 The
distribution of nodes which carry degree $k$ is denoted by $P(k)$.
This means we now look at the set of all observed frequencies
of symbol sequences, and consider the probability distribution
of these frequencies. For example, if all
symbol sequence probabilities are the same, as
for example for uncorrelated random sequences of infinite length, then $P(k)$
corresponds to a sharply peaked delta distribution.
The quantity  $P(k)$ is called {\it the degree distribution}. 
It characterizes
the network globally and
classifies it to be a scale-free network if $P(k)$
has power law tails,
\begin{eqnarray}
P(k)\sim k^{-\gamma}.
\label{eq073}
\end{eqnarray}
 $\gamma$ is the power law exponent expressing the
scale-free nature of the
network and it is typically in the range $2 <\gamma <3$,
although in some cases $\gamma$
may lie outside this interval.

\par
Apart from the degree distribution,
one of the most important variables in the theory of complex networks
is the local clustering
coefficient $c_n$ around the node $n$, which describes
the local structure
of the network around that specific node.
The local clustering coefficient
is defined as:
\begin{eqnarray}
c_n=\frac{\sum_{i,j}b_{ni}b_{ij}b_{jn}}{\sum_{i \neq  j}b_{ni}b_{jn}}
\label{eq077}
\end{eqnarray}

In Eq. \ref{eq077} the numerator is related to the total
weighted number of closed triangles
originating from node $n$, while the denominator
gives the maximum number of possible
triangles originating on the same node \cite{grindrod:2002,saramaki:2007}.
Sometimes it is possible to find the functional form of the
clustering coefficient $c(k)$  of nodes having degree $k$.
This is an important property of the network and indicates
an underlying  hierarchical structure \cite{ravasz:2003}.
For hierarchical networks a power law form is achieved
\begin{eqnarray}
c(k)\sim k^{-b }
\label{eq071}
\end{eqnarray}
where the exponent $b $ takes a positive value for hierarchical
networks, while it is constant for random uncorrelated networks and
for scale free networks. In many natural networks $b \sim 1$
\cite{ravasz:2003}. In general it is difficult to find such a
relation. It is important here to make the distinction between
$c(k)$, which is the functional form of the clustering coefficient
as a function of the degree $k$, and $c_i$ which is the clustering
coefficient of node $i$.

\par
The global clustering coefficient $c(V)$, defined as
the average of the local clustering
ones, characterises globally the connectivity in the network
and in general depends on the size $V$ of the network.
\begin{eqnarray}
c(V)=<c_i>=\frac{1}{V}\sum_{i=1}^V c_i.
\label{eq08}
\end{eqnarray}

For many real systems $c(V)$ is independent of $V$.
In particular,
the global clustering coefficient in  random uncorrelated
networks decreases as \cite{watts:1998}
\begin{eqnarray}
c(V)\sim V^{-1}.
\label{eq09}
\end{eqnarray}
In the case of scale-free, highly clustered and complex
networks Eq. \ref{eq09} changes to
\begin{eqnarray}
c(V)\sim V^{-\nu}.
\label{eq10}
\end{eqnarray}
\par The distribution of clustering coefficients $P(c)$
takes a power law form in scale free networks,
\begin{eqnarray}
P(c)\sim c^{-\beta}.
\label{eq11}
\end{eqnarray}
For random,
uncorrelated networks, it was shown by Watts and Strogatz  that
the local clustering coefficients have an exponential type of distribution
\cite{watts:1998,ravasz:2003}.

\par In view of the presence of self-loops in genomic sequences, their
contributions in the node degrees and the clustering coefficients
need to be commented on.
In the numerator
of Eq. \ref{eq077} the presence of the term $b_{kk}b_{kk}b_{kk}$ might
seem strange in social
networks but in the representation of symbolic sequence it
represents the phenomenon
of repeats, i.e. the repetition of the same string a number
of times in the sequence. If
the node $j$ represents the string $j\equiv [J_1,J_2, \cdots J_s]$,
where $J_i$ are symbols,
then the term
$b_{jj}b_{jj}b_{jj}$ denotes the presence of string
$j\otimes j\otimes j\otimes j $ in the
sequence.
Repetitions are very frequent in genomic sequences,
in particular for primates.
In the human genome one sequence repeat alone (the ALU-sequence) comprises
approximately 11.5\% of the human genome,  while the total repeat 
content reaches 35\% of the human DNA.

\section{Network Properties of DNA Sequences and of Intermittent Maps}
\label{sec3}

\subsection{DNA sequences}
\label{genomic}
In this section we first apply the network approach to genomic sequences,
following the ideas
described in the previous section. As working examples we use
chromosomes 10, 14 and 20 from the human genome.
\par In natural sequences such as in DNA  most often $b_{ij}
\ne b_{ji}$. In genomic sequences the two strands of the helix
 have complimentary structure. Let us call the two strands $C1$
and $C2$. This means that if a nucleotide A is found in a certain position
in $C1$ a nucleotide $T$ will be found in the sequence $C2$ in the same
position. Similarly, $T$ is the compliment of $A$, $C$ is the
compliment of $G$ and $G$ is the compliment of $C$.
Consider e.g.\ the string $S_1=[AGGT]$ followed
by $S' _1=[CGTT]$ both found in strand $C1$. Then in
strand $C2$, the following strings will
be found: $S_2=[TCCA]$ and $S' _2=[GCAA]$. Thus if we denote
by $\>\>\>\tilde{ } \>\>\>$ the complimentary strings and strands,
we have the following relation for the weighting matrices,
\begin{eqnarray}
b_{ij}(s)=\tilde b_{\tilde i\tilde j}(s)
\label{eq12}
\end{eqnarray}
It is then sufficient to compute the network characteristics of one
of the two strands and to mirror its properties to the other strand
according to Eq. \ref{eq12}.

\par
\begin{figure}
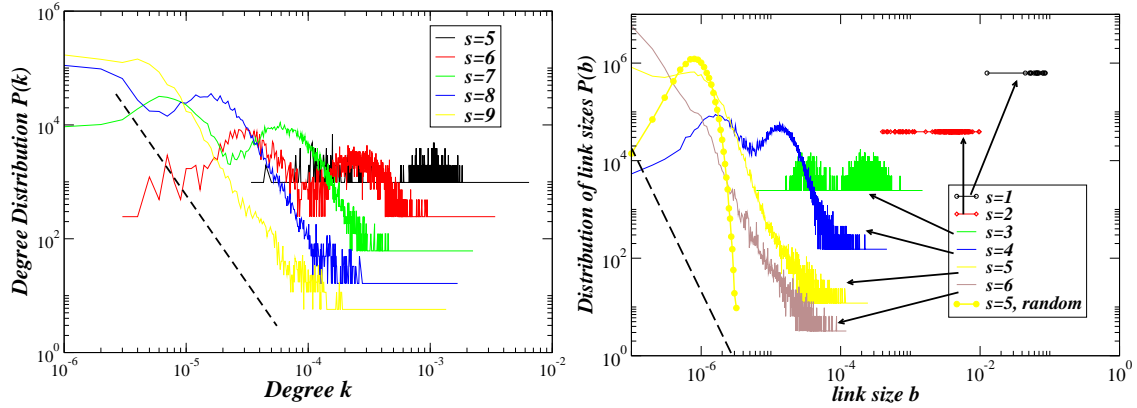

\includegraphics[clip,width=0.45\textwidth,angle=0]{./fig02a.eps}
\includegraphics[clip,width=0.45\textwidth,angle=0]{./fig02b.eps}\\

\caption{\label{fig:02}{} (Colour online)
Distribution functions related to the network derived from
 the Human chromosome 20.
a)The Degree Distribution $P(k)$ of the network of strings with various sizes $s$.
 The dashed line corresponds to a power law decay with exponent
$\gamma \sim -3$.
b) The Distribution $ P(b)$ of link weights $b_{ij}$ between nodes.
The dashed line corresponds to an exact power law with exponent $\gamma_1 = -3$.
The yellow bullets correspond to a random and uncorrelated sequence with $s=5$.
Strings of various sizes s
are plotted with different colours as indicated in the figure.
}
\end{figure}

 In Fig. \ref{fig:02}a the degree distribution
of the symbolic network
characterising the chromosome 20 genomic sequence of Homo sapiens
is presented.
Strings of different sizes were considered, up to $s=9$.
In the $x-$axis the degree $k$ characterising the total link strength carried by
a node is plotted, normalised with the total number of (weighted) links. This
normalisation is needed because the total number of links is a decreasing
function of the length $L$ of the symbol sequence. The $y-$axis shows the
distribution of nodes of degree $k$.  For comparison, the dashed line represents a pure power law
distribution with exponent $\gamma = -3$.

\par In Fig. \ref{fig:02}b the distribution of individual
link sizes (weights $b_{ij}$) is plotted independently of the node to which they
belong. String sizes $s=1-6$ are shown, taken also for the
human chromosome 20. Longer string sizes are not possible to
investigate due to computational limitations, since the size of the
matrix $\mathbf{b}$ grows exponentially with $s$.
The observed form of the $P(b)$
distribution is very similar to that of $P(k)$ in
Fig. \ref{fig:02}a. This is not unexpected since the
values in the latter figure represent cumulative  link weights
originating from one node. Again the dashed line corresponds
to power law behaviour with exponent $\gamma _1 = -3$.
The two exponents may not be exactly identical, due partly to
stochasticity and partly to  the fact that the degree is a sum over a finite
number of link sizes (over a node). If the number of
links on a node were infinite then
the two distributions would posses exactly the same
exponent $\gamma \equiv \gamma_1$.  For comparison,
the $P(b)$ distribution calculated from a random and
uncorrelated symbol sequence of the same size as chromosome 20
is plotted with yellow bullets.
The segmentation was done with $s=5$. In contrast to the
genomic data, the random symbol sequence shows
a hump around the mean value $5\times 10^{-7}$ and then drops
abruptly (step-like), as is expected for finite,
uncorrelated random sequences.
\par Note that for the case of symbol sequences the degree of a node
coincides with the frequency of appearance of the particular
string of length $s$.
For $s=1$ (one-letter words) there are
only 4 configurations and all of them have
similar frequency. That results in a narrow range
distribution with little structure.  
For $s=2$ (two-letter words) a first
appearance of two maxima is observed,
which correspond to the presence of multiple
$T$ and $A$ in the sequence. The minimum values
correspond to the infrequent presence of the
complex $GC$ in the system, which is known to
be related to the presence of functional units
called {\it promoters}. For $2<s<6$ the presence
of a larger number of strings/nodes in the network
smoothes the two well-pronounced maxima into
a two-humped distribution. Again, the two maxima
correspond to the presence of multiple $A$ and
$T$ strings, while the minimum is again corresponding
to the complexes of $GC$ and $CG$ followed by
one of the other four bps. For $s>5$ a power law
degree distribution establishes gradually,
which indicates the scale free character of this symbolic network.

\par
\begin{figure}
\includegraphics[clip,width=0.6\textwidth,angle=0]{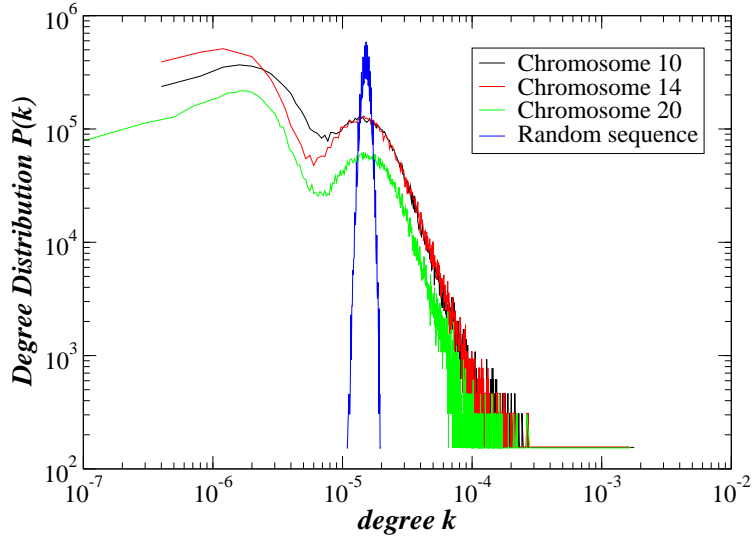}

\caption{\label{fig:03}{} (Colour online)
 The degree distributions for 3 different human chromosomes.
The degree distribution of a random symbol sequence of the
same size is also shown (humped distribution).}
\end{figure}

 For comparison, the degree distributions as computed for
 human chromosomes 10, 14 and 20 are plotted together
 in Fig. \ref{fig:03}.
The degree distributions of the three chromosomes
are qualitatively similar, which may point to a
universal type of scaling.
In the same figure the degree distribution
of a random sequence of the same size as chromosome 20 is
plotted. The random distribution is single-humped and is symmetric
around its mean value, as expected for random uncorrelated
sequences. Clearly, for infinitely long random sequences
one expects convergence to a $\delta$-function, whereas
for genomic sequences the distribution is
much broader.

\par To further explore the network connectivity we compute
the size distribution of clustering coefficients, throughout
the network. Due to computer memory limitations only strings of
size $s\le 6$ can be computed. To suppress fluctuations, the
cumulative size distribution $P_{cum}(c)$ is calculated as
\begin{eqnarray}
\label{eq13}
P_{cum}(C)=\int_{C}^{\infty}P(c)dc .
\end{eqnarray}
For power law distributions of the form \ref{eq11}, the
cumulative size distribution also follows a similar
power law, as
\begin{eqnarray}
\label{eq14}
P_{cum}(C)\sim\int_{C}^{\infty}c^{-\beta}dc\sim C^{-\beta +1} .
\end{eqnarray}

\begin{figure}
\includegraphics[clip,width=0.6\textwidth,angle=0]{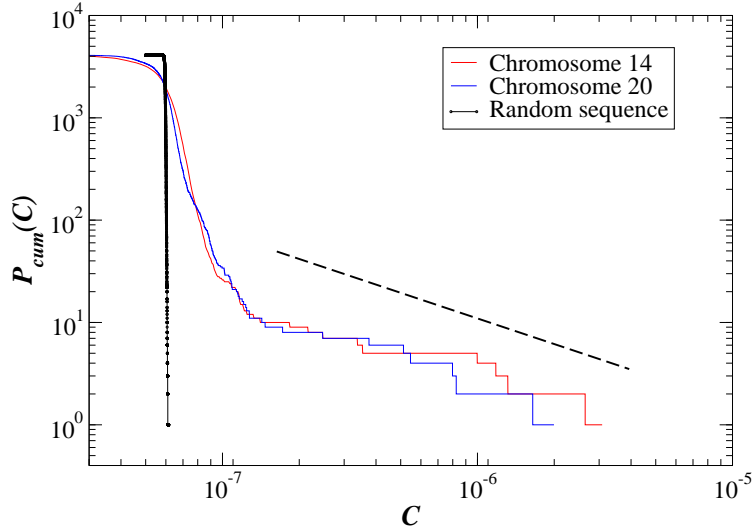}

\caption{\label{fig:04}{} (Colour online)
 The cumulative distribution of clustering coefficients $P_{cum}(c)$ is plotted
as a function of the size $c$ for human chromosomes 14 (red line),
20 (blue line) and
a random sequence (black line) of equal size with chromosome 20.
String size is $s=6$. For comparison the dashed line presents a pure
power law decay with exponent $-1.7$.}
\end{figure}
In Fig. \ref{fig:04} the cumulative clustering coefficient distribution is
plotted as a function of the coefficient size $C$. Data from chromosomes 20
and 14 are plotted together with data taken from an artificial random symbol
sequence whose symbol frequencies are the same as in chromosome 20. In a
double logarithmic scale the genomic cumulative distributions exhibit an
almost linear regime for large sizes, indicating the presence of a power
law. This behavior becomes more prominent as the string size
increases. In comparison, the data from the 
large-length random sequence has an abrupt,
almost step-like decay, indicating a very sharply
 peaked  Gaussian ($\delta$-like) distribution,  whose
cumulative distribution function is very close to a step-like function.

\subsection{Polynomial map}
\label{poly}

\par 

Methods to construct
networks from maps or a given time series have been previously addressed
in refs. \cite{nicolis:2005, net1, net2, robledo}, using as a particular examples the tent map,
the cusp map, or the logistic map.
In \cite{nicolis:2005}, the phase space of the maps is segmented
into a number of cells and each cell corresponds to a node of the
network. The connectivity matrix is then defined by the frequency
of transitions between the different cells/nodes of the network.

\par The current approach is inspired by \cite{nicolis:2005}, but the
transition from the map to the network is achieved using
symbolic sequence generated by the map.
In other words, the map dynamics is first mirrored on
a symbol sequence as explained in sec. \ref{symbol-sequences} and then the network is
constructed from the symbol sequence as discussed in sec. \ref{networks}.
The choice of the polynomial map, mentioned briefly in sec. \ref{polynomial-map},
 is based on its intermittent behaviour
and its capacity to give rise to time series (and corresponding symbol
sequences) with long range features, as opposed to the dynamics of the
logistic map and other non-intermittent maps giving rise to 
nearly uncorrelated
behavior.

\par
\begin{figure}
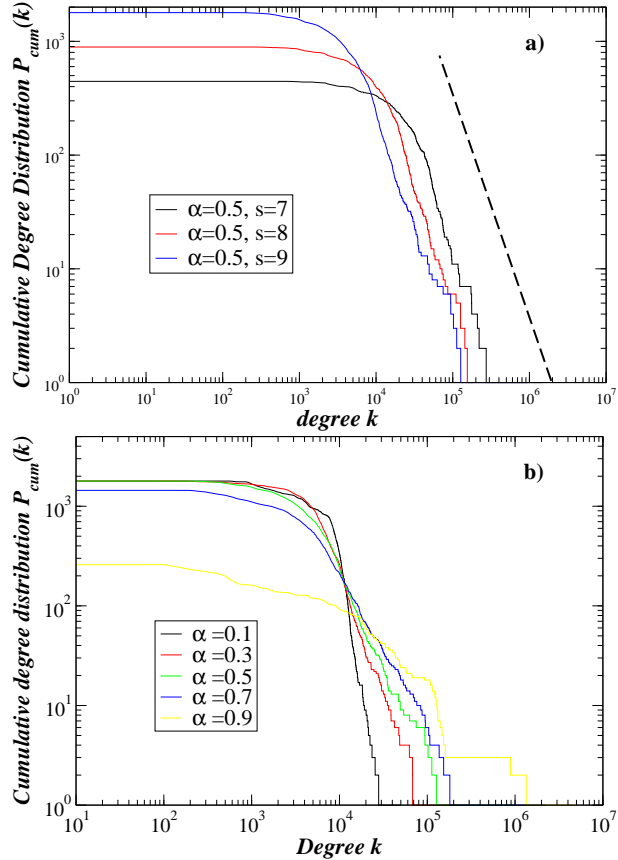

\includegraphics[clip,width=0.49\textwidth,angle=0]{./fig05a.eps}\\
\includegraphics[clip,width=0.49\textwidth,angle=0]{./fig05b.eps}


\caption{\label{fig:05}{} (Colour online)
Network symbol representation of the polynomial map.
a) Cumulative degree distribution for parameter value
$\alpha =0.5$ and various string sizes. The dashed line
corresponds to an exact power law with exponent $\gamma =-3$.
b) Cumulative degree distribution for string size
$s=9$ and various parameter values.
}
\end{figure}

 In Fig. \ref{fig:05}a the cumulative degree distribution for
the polynomial map  is shown for
various values of string sizes $s$ and  parameter value $\alpha =0.5$.
The sequence size was chosen as $L=4.3\cdot 10^7$, of comparable size as
chromosome 20. In our plots we have chosen
the cumulative degree distribution rather than the probability density function to somewhat smoothen out fluctuations.
The frequency of appearance
of each nucleotide is chosen as in Eq. \ref{eq03} and corresponds
to those of human chromosome 20. For this particular parameter value,
all string sizes point towards the same exponent
$\gamma (a=0.5)\sim 3$.
Note that the number of allowed string
configurations $V$ generated by the polynomial map is far less than the
number of strings observed in human genomic sequences. As an example
we note that
$V_{DNA}(s=9)=244925 < 4^9=262144$, while $V_{poly}(s=9)=1790$.

\par In Fig. \ref{fig:05}b the cumulative degree distribution for
the polynomial map  is shown for
various values of the  parameter value $\alpha $ and string sizes $s=9$.
It is obvious that the exponent $\gamma$ is a decreasing function of
the parameter $\alpha$. By appropriate choice of the value of $\alpha$
we can achieve the same power law exponent as the one observed
in the human chromosome.
 On the other hand, the number of configurations generated
by the polynomial map ($\sim 1700$) is far less than observed
in genomic sequences ($\sim 250 000$ in chromosome 20). This difference
is non-trivial, it covers 2 orders of magnitude.
To achieve the diversity of the string configurations
together with the degree distribution scaling observed in genomic sequences,
a diffusive coupling is introduced in the next section
between a large number of polynomial maps (considered as "units").
This will create a large variety of string configurations
together with similar exponents as for genomic networks.

\section{Network Properties of Coupled Polynomial Maps}
\label{sec4}
Coupled Map Lattices (CML) have been extensively used for the
modelling of many physical systems which involve interactions
between many spatially separated
constituents. A lot of emphasis of research activity
has been put on spatio-temporal chaos and
synchronization phenomena arising in CMLs
\cite{kaneko:1984,kapral:1985,kaneko:1993,brannstrom:2005,lin:2002,schmitzer:2009,provata-beck:2011}.
\par For the coupling of polynomial maps, in the present study,
a simple, 1-dimensional chain arrangement with periodic boundary conditions
is assumed. The periodic boundary conditions are chosen simply for
convenience and they do not affect, qualitatively or quantitatively,
the results in the limit of very long chains, as considered
here.
\par Our linear chain arrangement consists of $L=10^8$ polynomial
maps coupled to their nearest neighbours with a coupling constant
$r$. The dynamics is
\begin{eqnarray}
x_{n+1}^i= \left\{
\begin{array}{l l l}
(1-r)x_{n}^i(1+2^{\alpha}(x_{n}^i)^{\alpha })+\frac{1}{2}r
\left[ x_{n}^{i+1}(1+2^{\alpha}(x_{n}^{i+1})^{\alpha })
     + x_{n}^{i-1}(1+2^{\alpha}(x_{n}^{i-1})^{\alpha }) \right] &\quad {\rm if} & x_{n}\le 0.5 \\
(1-r)2x_{n}+\frac{1}{2}r\left[ 2x_n^{i+1}+2x_n^{i-1} \right] &\quad {\rm if} & x_{n}>0.5 \\
\end{array}
\right.
\label{eq021}
\end{eqnarray}

The values $x^i_n$ are taken modulo 1 for all $n$, as in Eq. \ref{eq01}.
 The index $i=1,2 \cdots $ runs over all local maps,
while $n=1,2,\cdots $ is a temporal index.
Random initial conditions are chosen for each map.
The parameter value is chosen as $\alpha =0.5$ and  the
number of iterations in our simulation
is $T=5000$, sufficiently high for the
maps to enter their dynamic equilibrium regime.  At $T=5000$ the
state of each map is recorded and a transformation to a symbol
sequence is performed using Eq. \ref{eq03}, with the same
1-point symbol sequences as for the chromosome data.
At the final stage the
symbol sequence is divided into strings of size $s$ and the
corresponding network connectivity matrix $\mathbf{b}$ is constructed
according to the method described in Sec. \ref{networks}.

\begin{figure}
\includegraphics[clip,width=0.6\textwidth,angle=0]{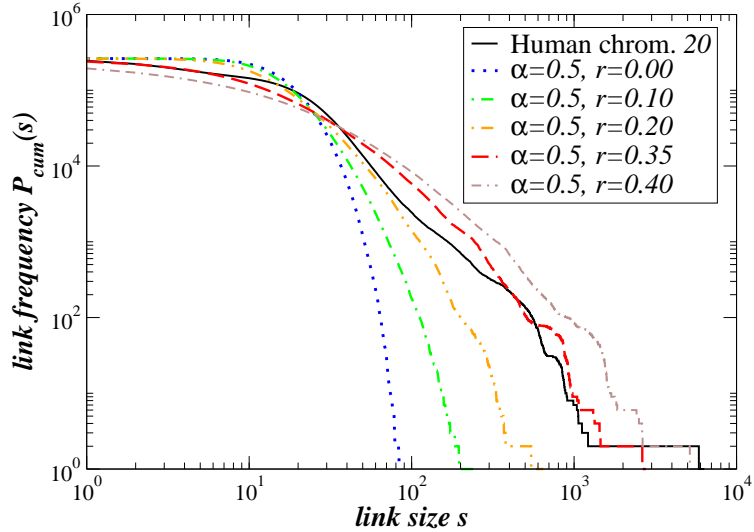}

\caption{\label{fig:06}{} (Colour online)
Coupled polynomial maps on a linear chain.
 The Cumulative Degree Distribution $P_{cum}(k)$
of the network is plotted for
various values of the coupling constant $r$.
For comparison the corresponding data for chromosome 20 are plotted
with the black solid line.
Parameter values are $\alpha =0.5,\>\> T=5000, \>\> L=4.3 \> 10^7,
\>\> s=9 $.
The symbol frequencies were chosen as in Eq. \ref{eq03}.
Various values of the coupling constant $r $
are plotted, as indicated in the legend.
}
\end{figure}

\begin{figure}
\includegraphics[clip,width=0.6\textwidth,angle=0]{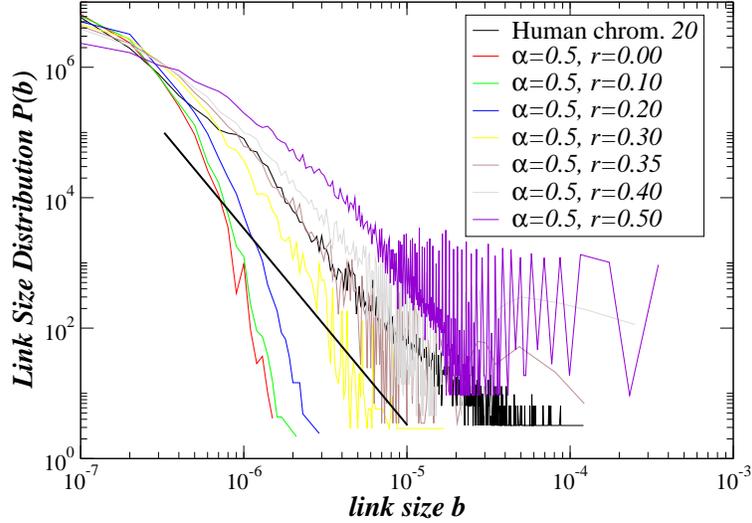}

\caption{\label{fig:07}{} (Colour online)
Coupled polynomial maps on a linear chain:
The Link Size Distributions $ P(b)$ of
link weights $b_{ij}$ between all the nodes
is plotted for various values of the coupling $r$.
The data of chromosome 20 are represented by the
black solid line.
Parameter values are $\alpha =0.5,\>\> T=5000, \>\> L=4.3 \cdot 10^7,
\>\> s=6 $.
The symbol frequencies were chosen as in Eq. \ref{eq03}.
Results for various coupling rates $r$ are shown. The solid
straight line represents an exact power law with exponent -3.}

\end{figure}

\begin{figure}
\includegraphics[clip,width=0.6\textwidth,angle=0]{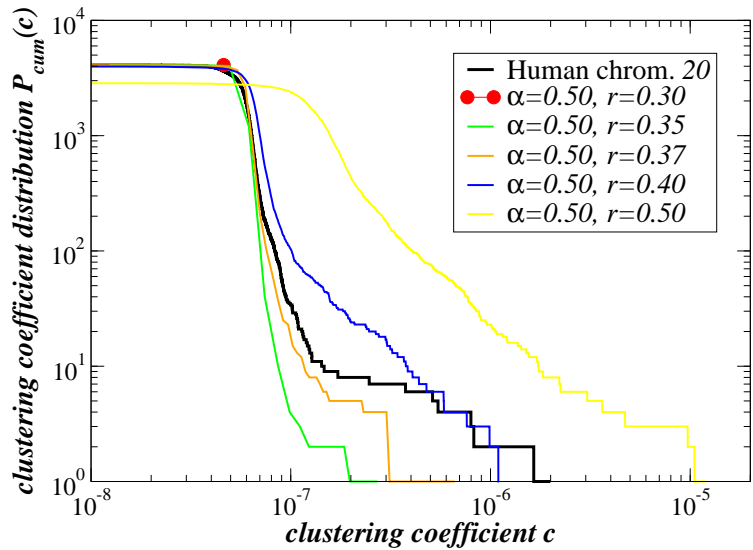}

\caption{\label{fig:08}{} (Colour online)
The  cumulative distribution of clustering coefficients $P_{cum}(c)$ is plotted
for various values of the
coupling $r$. For comparison the data of chromosome
20 are plotted as a black solid line.
All parameters (including nucleotide
frequencies) are chosen as in
Fig. \ref{fig:07}.
}
\end{figure}

\par In Figs. \ref{fig:06}, \ref{fig:07} and \ref{fig:08}
the  cumulative degree distribution $P_{cum}(k)$, the  link size distribution $P(b)$,
and the cumulative distribution of
clustering coefficients $P_{cum}(c)$
are plotted for
various values of the coupling constant $r$.
For comparison, the corresponding data for chromosome 20 are also plotted
in each figure.
\par For the calculation of the degree distribution  window
size $s=9$ is used. In Fig. \ref{fig:06} the cumulative
distribution is plotted.
Comparison of the different curves reveals
that the coupled polynomial maps with parameter $\alpha =0.5$
and coupling constant of the order of $r\sim 0.35$ assimilate
relatively well the sequence structure of Chromosome 20.
\par For the calculation of the link size distribution strings
of size $s=6$ were employed. This is because transition matrices
of size $4^s \times 4^s$ need to be considered which are very
demanding in computer memory. The results for $s=6$ are plotted
in Fig. \ref{fig:07} both for Human Chromosome 20 (black solid
line) and coupled polynomial maps with parameter $\alpha =0.5$
and various values of the coupling constant $r$. Again the best fit
is observed for $r\sim 0.35$ despite the fact of using a different
(smaller) string size $s$ for the calculations.
This shows that the
similarities between the statistics of chromosomes
and coupled polynomial maps are
robust to variations in the size of window used in the creation
of the network, provided that $s$ is not too small ($s > 5$).
The observed power law exponent is again of the order $ \sim 3$, as
represented by the straight line in the double logarithmic scale
in Fig. \ref{fig:07}.
\par Finally, the distributions of  clustering coefficients are
 presented in Fig. \ref{fig:08}. Again,  the
genomic data (Chromosome 20) are plotted together with sequences
resulting from coupled
polynomial maps with $\alpha =0.5$ and various coupling rates
$r$. The results are consistent with the previous findings.
While for small values of $r$ the distribution of clustering
coefficients drops abruptly as in random sequences, as $r$ grows
the distribution develops a long tail which approaches the tails
of DNA sequences around the coupling values $r\sim 0.35-0.40$.

\par  We notice that uncoupled polynomial
maps can not well represent the complexity of DNA sequences, although they
are known to produce intermittency with long range correlations.
On the other hand, a medium size coupling
between polynomial
maps is able to create the appropriate correlations and to resemble
the structure of DNA in many levels
of complexity.  From the last three figures one can see that a coupling constant
of the order of $r=0.35$ is enough to adjust the power law exponents to
values close to the ones observed in genomic sequences.
The need for a  coupling between neighboring units to properly
assimilate DNA sequences demonstrates the presence of local interactions
between the adjacent nucleotide strings  which create
the correlated, mosaic structure of the genome.
\par Similar conclusions are obtained from the network analysis of
other chromosomes. Just slight variations
in the values of the exponents and the
necessary coupling constants are noted due to the
difference in the symbol frequencies in the chromosomes and
due to stochastic effects.

\par From our analysis we also see that quite generally coupled polynomial
maps give rise to complex small-world networks, via the corresponding
symbol sequences and transition matrix, while the network exponents can
be adjusted by varying the coupling constant $r$.

\section{Conclusions}
\label{sec5}

The dynamics of coupled intermittent maps was used to model the
correlated structure of genomic sequences via a network approach.
The weighted network approach to symbolic sequence was first introduced 
and applied to genomic and random, uncorrelated sequences and then 
compared with the corresponding statistics of coupled intermittent maps.
For the modelling the use of intermittent
maps appears to be necessary in order to retrieve the scaling
properties observed in the primary structure of DNA.
It was first shown that although the dynamics of single intermittent maps produce
long range correlated symbolic sequences, with a variety of power law exponents
depending on the choice of the parameters,
they do not produce the diversity of genomic strings observed in DNA
sequences.
 To overcome this limitation coupled map lattices were considered,
with diffusive coupling between neighboring units
on a 1-dimensional lattice.
It was shown that a medium size coupling between neighboring polynomial
maps  is sufficient to produce a) power law exponents comparable with the ones obtained
from genomic data and b) a statistical
distribution of string frequencies similar to real DNA sequences.
Our results are consistent with the known
existence of complicated patterns of correlations between adjacent segments in DNA.
\par The reported results concern the primary structure of human
chromosomes. The  network method can be applied to any genomic sequence provided
it is long enough to assure reasonable statistics. It would be of
great interest to  study further
classes of organisms with this method and explore the range of values of the
network exponents for different organisms.  Additionally, the proposed
network approach to symbol sequences may be used to construct quite generally networks
from any symbol sequence (natural, experimental or artificial) and to test
for scaling characteristics.

\end{document}